\documentclass[aps,prl,twocolumn,superscriptaddress,showpacs,showkeys]{revtex4}
\usepackage{amsmath,amssymb,epsfig}

\begin{document}

\preprint{NSF-ITP-02-67}

\title{Real-time determination of free energy and losses in optical
absorbing media }


\author{C. Broadbent}
\affiliation{Department of Physics, Brigham Young University,
Provo, Utah, 84601}
\author{G. Hovhannisyan}
\affiliation{Department of Mathematics, Brigham Young University,
Provo, Utah 84601 }
\author{J. Peatross}
\affiliation{Department of Physics, Brigham Young University,
Provo, Utah, 84601}
\author{M. Clayton}
\affiliation{Department of Mathematics, Penn State University,
University Park, State College, Pennsylvania 16802}
\author{S. Glasgow}
\affiliation{Department of Mathematics, Brigham Young University,
Provo, Utah 84601 }



\date{\today}

\begin{abstract}
We introduce notions of free energy and loss in linear, absorbing
dielectric media which are relevant to the regime in which the
macroscopic Maxwell equations are themselves relevant. As such we
solve a problem eluded to by Landau and Lifshitz \cite{LL84} in
1958, and later considered explicitly by Barash and Ginzburg
\cite{BG76}, and Oughtsun and Sherman \cite{OS94}. As such we
provide physically-relevant real-time notions of "energy" and
"loss" in all analogous linear dissipative systems.

\end{abstract}


\maketitle


In a previous publication we showed that in many instances fast
and slow light \cite{GM70,CW82,HHDB99} is a manifestation of a
dielectric interacting differently with the early parts of an
electromagnetic pulse than with its later parts \cite{JOSAA}. For
example, slow light in a passive linear dielectric typically
corresponds to energy in the leading part of the pulse being
preferentially stored by the medium and then being largely
returned to the pulse's (backward) tail. Quantifying the extent to
which this process is possible was a primary motivation for this
work.

Application of the principles presented in this letter will in
no-wise supersede the group velocity description of such
phenomena. The present development compliments other such analysis
by attaching precise notions of reversibility and irreversibility
to the medium's storage of the pulse energy. As such it also
solves a long outstanding problem regarding the real-time meaning
of "energy" and "loss" in linear dissipative systems.

From Poynting's energy conservation theorem (Eq. \ref{Eq:Poyn})
the total energy density in a dielectric at time $t$, $u(t)$, is
the sum (Eq. \ref{Eq:utot}) of the field energy $u_{field}(t)$
(Eq. \ref{Eq:ufield}) and the medium-field interaction energy
$u_{int.}(t)$ (Eq. \ref{Eq:uint}):

\begin{align}
\nabla \cdot {\bf S }+ \frac{\partial u}{\partial t} &= 0 \label{Eq:Poyn}\\
u(t) &= u_{field}(t)+u_{int.}(t) \label{Eq:utot}\\
u_{field}(t) &= \frac{E^2}{2} + \frac{H^2}{2}
\label{Eq:ufield}\\
u_{int.}(t) &= \int\limits_{-\infty}^t E(\tau)\dot{P}(\tau) \,
d\tau .\label{Eq:uint}
\end{align}

Here we restrict to isotropic, temporally dispersive media in
which a scalar analysis suffices, and in which one may safely
suppress reference to the spatial coordinate ${\bf x}$. In
addition we restrict to linear media. Consequently we restrict to
the case in which the medium's response to an applied field ${\bf
E}$ is completely determined by a scalar, point-wise defined
susceptibility $\chi=\chi(\omega,{\bf x})$. For a basic discussion
of properties of the medium-field interaction energy $u_{int.}(t)$
in anisotropic media, see \cite{GWP01}. Since $u_{int.}(t)$
quantifies the net work the field has done against the polarized
medium in the course of creating the current system state, in the
following we will also call $u_{int.}(t)$ the (medium) internal
energy. In the sequel, notions of work done, and the ability to do
work in the future, will become pivotal in establishing an
unambiguous, dynamically relevant notion of energy allocation in
dielectric media.

Landau and Lifshitz interpreted $u_{int.}(+\infty)$ as the energy
that is eventually lost to and dissipated by the medium over the
course of the medium-field interaction \cite{LL84}.  This
asymptotic quantity only depends on the medium susceptibility
$\chi$ and the evolution of the electric field $E$, not on the
particulars of any microscopic model giving rise to $\chi$. This
suggests the possibility of establishing a real-time, model
independent notion of loss. Barash and Ginsburg considered this
question \cite{BG76}.  They concluded, however, that a real-time
determination of losses is impossible without a microscopic model,
i.e. that a knowledge of the macroscopic susceptibility $\chi$ is
insufficient. Having concluded that a model independent notion of
loss is meaningless, they calculated (a certain notion of) loss
for specific models. In particular they generalized the work of
Loudon \cite{L70} concerning single Lorentz oscillator media by
making a straightforward extension of his notion of losses to
multiple Lorentz oscillators. Loudon's and, so, Barash and
Ginsburg's notion of losses amounts to identifying those terms in
(a certain expression of) the internal energy which explicitly
depend on phenomenological damping parameters. More convincingly,
their determination of the "energy" in the medium amounts to
summing the kinetic and potential energies of the individual
oscillators. We call this notion of energy the \emph{polarization}
energy, $u_{polarization}(t)$.

We initially followed a certain extension of this reasoning. We
hoped to establish a unique map from an arbitrary susceptibility
$\chi(\omega)$ to an oscillator representation, which in turn
would establish a unique (polarization) energy, as well as losses
via Loudon's identification procedure. Unfortunately, we found
that generically a susceptibility can be mapped to many different
microscopic models. This property is perfectly analogous to the
one by which it is possible for distinct LRC circuits to yield
identical impedances. However, we then conjectured that all such
equivalent representations might yield identical \emph{values} for
the polarization energy. Unsurprisingly, one finds that this is
not the case dynamically. (Although asymptotically all such
representations must obviously agree:
$u_{polarization}(+\infty)=0$, and the losses are given, then, by
$u_{int}(+\infty)$, which, as mentioned, is completely determined
by $\chi$.)

\begin{figure}
    \includegraphics{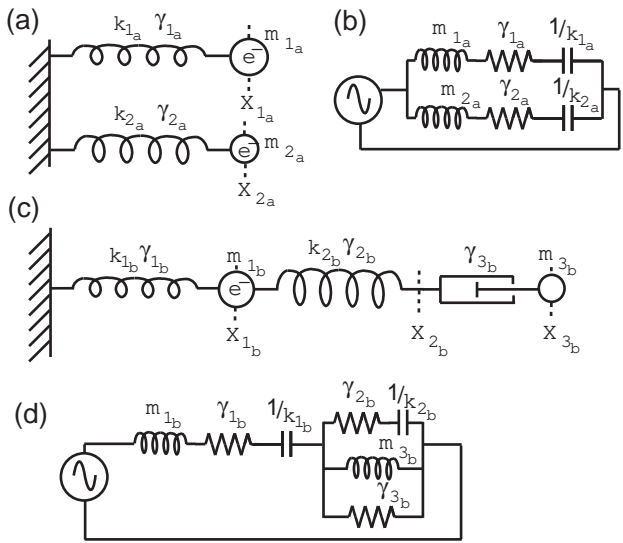}
    \caption{Macroscopic equivalent spring-mass systems for a double
    Lorentz oscillator and their LRC circuit analogs. (a) The two
    oscillator representation.  (b) The LRC circuit analog of (a).
    (c) The coupled oscillator representation.  (d) The circuit analog of (c).\label{fig:sprgms}}
\end{figure}

\begin{figure}
    \includegraphics{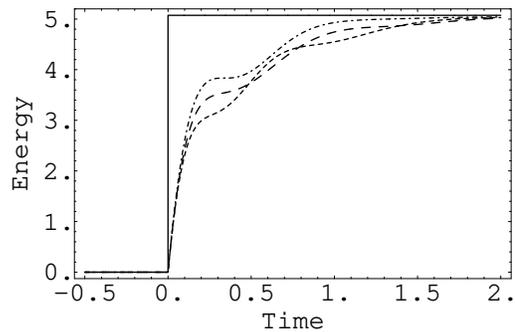}
    \caption{The instantaneous losses of a
    two oscillator (long dash), and a coupled oscillator (short dash) model of a
    double Lorentz oscillator susceptibility
    responding to an instigating delta function E-field at time
    $t=0$.  The interaction energy
    $u_{int.}[E](t)$ (solid), and the irreversible energy $u_{irrev.}(t)$ (dot-dash) are also shown.\label{fig:free}}
\end{figure}

Figures \ref{fig:sprgms} and \ref{fig:free} demonstrate this
ambiguity. Given a double Lorentz oscillator susceptibility
$\chi(\omega) = \sum\nolimits_{n=1,2}
\omega_{p_n}^2/(\omega_n^2-i\gamma_n\omega-\omega^2)$ we examine
two different microscopic models for the same susceptibility.  The
first model $\chi_{a}(\omega)$, is given by the explicit structure
of $\chi(\omega)$ as just written. It corresponds to two different
masses of equal charge on two different springs, each spring
having different restoring and damping parameters. The total
polarization is given by the net displacement of both charged
masses from their equilibrium positions:
$P=\textup{X}_{1_a}+\textup{X}_{2_a}=\chi_a E$. See
FIG.~\ref{fig:sprgms}(a).

We use the tangency algorithm~\cite{GMCHB}, a transfer function
preserving algorithm intended for model reduction in LRC circuits,
to construct the second microscopic model $\chi_b(\omega)$.
Mechanically this model corresponds to coupled oscillators. See
FIG.~\ref{fig:sprgms}(c). The susceptibility for this microscopic
model is given in Eq. (\ref{Eq:chib}) (where we have mapped
$\omega$ to $i s$ for convenience). In this model the total
polarization is given by the displacement of the charged mass from
its equilibrium position $P=\textup{X}_{1_b}=\chi_b E$. It can be
shown that the coefficients $k_{n_b},\gamma_{n_b},m_{n_b}$ can be
found such that $\chi_a(\omega)=\chi_b(\omega)$ for all
frequencies $\omega$.
\begin{eqnarray}
\chi_b(s) = \frac{1}{k_{1_b}+\gamma_{1_b}s+m_{1_b}s
^2+\frac{1}{\frac{1}{k_{2_b}+\gamma_{2_b}s}+\frac{1}{\gamma_{3_b}s}+\frac{1}{m_{3_b}s^2}}}
\label{Eq:chib}
\end{eqnarray}

 By inserting the two representations into the internal energy $u_{int}(t) =
\int\nolimits_{-\infty}^{t} E(\tau)\dot{P}(\tau)\,d\tau$ one can
find the kinetic and potential energies associated with each
representation.  This polarization energy, the energy reckoned
instantaneously via the motion and position of the masses, will be
shown to be representation dependent. To demonstrate this, we
calculate the "microscopic" energetics associated with the
response of these macroscopically equivalent systems to an
impulse. That is we excite the two media, with susceptibility
models $\chi_a$ and $\chi_b$, via a delta function E-field at time
$t=0$, and then plot the losses for each representation. We assign
values to the parameters of the two oscillator, $\chi_a$
representation (the plasma frequencies, resonant frequencies, and
damping coefficients), and then find parameters for the coupled
oscillators', $\chi_b$ representation such that the
representations are macroscopically equivalent. Figure
\ref{fig:free} shows the evolution of the losses for the different
representations, these losses determined in the sense of Loudon,
Barash and Ginzburg. The corresponding differences between the
internal energy imparted to the media by the delta function
E-field (as shown by the piecewise constant curve in figure
\ref{fig:free}) and the plotted losses gives the polarization
energies for the two representations.

The qualitative features of Figure \ref{fig:free} can be
understood intuitively: the delta excitation of the medium
instantaneously creates the polarization energies, which, then,
decrease as the systems dissipate these energies. However, as
clearly seen in the figures, the polarization energies for each
microscopic representation differ significantly. Thus the
polarization energy does in fact depend on the microscopic model
ascribed to the susceptibility $\chi(\omega)$. Of course, since
Barash and Ginzburg specified no underlying macroscopic physical
principle in their determination of "energy", this result is just
a consequence of their (lack of a) definition of energy.
Consequently, we will argue that a model-dependent (e.g.
polarization) energy is irrelevant, both from the point of view of
the relevant physical principles that should be required of a
viable macroscopic description, as well as from that of the
practical considerations of the energy storage and return process
mentioned at the beginning of this article.

Before establishing relevant macroscopic principles of energy
allocation, we highlight the connection between mechanical and
electrical oscillators and the ambiguities associated with
representation in the context of the latter.  The perfect analogy
between electrical LRC circuits and linear passive dielectric
media (where polarization $P$, susceptibility $\chi$ and E-field
$E$ , translate to charge $Q$, derivative of admittance
$\partial_t A$, and electromotive force $\mathcal{E}$,
respectively), allows one to reinterpret the polarization energy
evolutions implied in Fig. \ref{fig:free} as the evolutions of the
energies contained in the dispersive elements (inductors and
capacitors) of admittance-equivalent circuits. In the case of
electrical circuits, the plotted losses correspond to the losses
in the dissipative elements of the two circuits, i.e. the losses
through the resistors. This time we interpret the discrepancies in
these curves as indicating that different LRC circuits with the
same admittance can, temporarily, lose energy to their resistors
at different rates. Figures \ref{fig:sprgms}(b)and (d) give LRC
circuits corresponding to the mechanical models shown in Figures
\ref{fig:sprgms}(a) and (c), respectively.

From the discussion above, one concludes that the polarization
energy and the associated losses depend intrinsically upon the
specific microscopic model giving rise to the susceptibility
$\chi$. However, these allocations cannot be relevant
macroscopically: for example, within the phenomenological
framework in which $\chi$ is introduced, the spatial and temporal
evolution of the various fields depend only upon this particular
piece of information, not upon its various representations. In
this circumstance in which the system dynamics is completely
determined by some piece of information (e.g. $\chi$), to say that
some other piece of information is important in order to establish
some particular notion of energy allocation is to admit that that
notion of energy allocation is irrelevant to those all-ready
specified dynamics. We establish a notion of energy allocation
that is determined uniquely by $\chi$ and, so, is relevant to
system dynamics. In particular it provides a precise notion of the
maximum possibility for the field to recover energy from the
medium in, for example, the borrow-return process mentioned at the
beginning of this article.

In this article, we introduce the \emph{irreversible energy}
density $u_{irrev.}(t)$. At any given time $t$, it is defined to
be the infemum of all possible [LL] asymptotic losses
$u_{int.}(+\infty)$, this extrema being realized over all possible
future evolutions of the electric field $E$, holding its past
evolution fixed. Temporarily using notation emphasizing that the
internal energy at time $t$ depends not only on this time, but
also on the history of the instigating electric field $E$ up until
that time, one writes

\begin{align}
u_{irrev.}[E](t)=u_{irrev.}[E_t^{-}](t)
&:=\underset{E_t^{+}}{\inf} u_{int.}[E_t^{-}+E_t^{+}](+\infty).
 \label{Eq:uirrev}
\end{align}

Here $E_t^{-}$ denotes the electric field time series $E(\tau)$
with its $t$-future ($\tau > t$) eliminated. Similarly, $E_t^{+}$
denotes an appended electric field time series $E(\tau)$ with its
$t$-past ($\tau < t$) eliminated. For a passive dielectric, this
infemum exists and, so, is unique \cite{AMEREM}. In particular,
and is shown later in Eq. (\ref{Eq:RI}), it does not depend upon
an explicit representation for $\chi$.

Almost tautologically, from definition (\ref{Eq:uirrev}), it
follows that $u_{irrev.}(t)$ can never decrease as time increases.
Thus, at any given time $t$, it quantifies a component of the
medium internal energy $u_{int.}(t)$ that will, under all
circumstances, remain in and be dissipated by the medium. That is
it specifies a component of the medium internal energy that cannot
under any circumstances be returned to the field. Moreover, since
it is defined in terms of an infemum, i.e. a greatest lower bound,
all notions of loss greater than this value are too pessimistic:
at any given time $t$, there always exists a future medium-field
interaction creating less eventual energy loss to the medium than
any value greater than that specified by $u_{irrev.}(t)$. This is
true regardless of how small this overestimation is. Consequently,
within the phenomenological, macroscopic framework in which $\chi$
dictates the system dynamics, this quantity uniquely records the
energetic irreversibility generated within this dissipative
system.

In Fig. \ref{fig:free} the dot-dashed curve specifies the
irreversible energy for the case considered, obviously valid for
either of the two $\chi$-equivalent physical systems. Note that in
the figure it is never exceeded by the losses in the sense of
Loudon, Barash and Ginzburg. Indeed one can prove that this
relationship must hold between the macroscopically relevant
irreversible energy and any notion of loss specified
microscopically. Equivalently one determines that energy (the
ability to do work) specified macroscopically cannot exceed any
such microscopic notion. Further, one concludes that the former is
almost always strictly less than the latter because of incoherence
among the system's microscopic, energy containing elements. The
former statement (the one regarding losses) can be obtained by
repeated application of the following theorem: (As in the
monotonicity of $u_{irrev.}(t)$, the theorem follows almost
tautologically from the definition.)

\begin{align}
u_{irrev.}[E;\chi_1+\chi_2](t) \geq
u_{irrev.}[E;\chi_1](t)+u_{irrev.}[E;\chi_2](t),
 \label{Eq:entropy}
\end{align}

with strict inequality holding almost always for nontrivial
$\chi_1$ and $\chi_2$. Eq. (\ref{Eq:entropy}) demonstrates that in
"additive" processes (i.e. generating media mixtures),
irreversibility is generated. By repeated (additive) subdivisions
of a macroscopically relevant susceptibility $\chi$ into pieces
$\chi_i$, one may obtain microscopic representations of the medium
response. If the elements are "simple" enough (to be quantified
later) $u_{irrev.}[E;\chi_i]$ will be equivalent to the
parameter-dependent notion of loss suggested by Loudon, and
Ginzburg and Barash. Indeed the notion of loss and energy
specified by Loudon happen to agree with the
macroscopically/phenomenologically relevant notion herein
introduced in the cases he considered-- single Lorentz oscillator
media. [This is not the case for (competing) multiple oscillator
media considered by Barash and Ginzburg, as demonstrated by figure
\ref{fig:free}.]

The difference between the current internal energy and the current
irreversible energy gives the \emph{reversible energy}
$u_{rev.}(t)$:

\begin{align}
u_{rev.}(t):= u_{int.}(t)-u_{irrev.}(t).
 \label{Eq:urev}
\end{align}

The reversible energy gives the least upper bound on the amount of
energy that the medium can relinquish after time $t$: any amount
greater than this value, no matter how small the discrepancy,
overestimates the ability of the medium's microscopic,
energy-containing components, say, to organize themselves and do
useful, macroscopic work. Obviously the difference between the
(piece-wise constant) internal energy plotted in Fig.
\ref{fig:free}, and the dot-dashed curve in that figure, gives
this dynamical notion of the possibility for the medium to do work
(against the field) for the example considered.

Using Eq. (\ref{Eq:uint}), one immediately shows that
$u_{int.}(t)$ is constant after the electric field ceases. From
definition (\ref{Eq:urev}), and the theorem regarding the
monotonicity of $u_{irrev.}(t)$, it follows that $u_{rev.}(t)$ can
never increase after such time, i.e. after the electric field
quits subsidizing its existence by doing work against the
polarized medium. We will show that $u_{rev.}(t)$ is never
negative.  Consequently, one sees that when the system becomes
energetically closed, the reversible energy behaves like a
dynamical system free energy (density), equivalently like a system
Lyapunov function (density). For this reason, and for the
microscopic consideration mentioned, in particular the
entropy-generation-like statement embodied in Eq.
(\ref{Eq:entropy}), we will also designate the reversible energy
as the (medium-field) \emph{free energy}.

We finish with a formula demonstrating how the macroscopic loss
(and so the free energy (via \ref{Eq:urev})), can be calculated.
(In particular how the irreversible energy plotted in figure
(\ref{fig:free}) can be generated.) This formula is obtained by
applying a variational principle to the definition
(\ref{Eq:uirrev}), and solving the resulting
\emph{Riemann-Hilbert} problem. One finds that, for passive,
causal dielectrics,

\begin{eqnarray}
\begin{aligned}
u_{irrev.}[E](t) &=\frac{\lambda}{2 \pi}
\int\limits_{-\infty}^{t}\left|\int\limits_{-\infty}^{+\infty}
\frac{-i\omega\chi(\omega)E_{\tau}(\omega)}{\phi_+(\omega)}e^{-i\omega
\tau} \, d\omega \right|^2 \, d\tau \label{Eq:RI}
\end{aligned}
\end{eqnarray}
where
\begin{eqnarray}
\begin{aligned}
\lambda &= \lim_{\omega \to \infty}
\frac{\textup{Im}[\chi(\omega)]}{\omega\left|\chi(\omega)\right|^2},\\
\phi_+(\omega) &= \lim_{\epsilon \to 0^+} \exp\left[\frac{-1}{2\pi
i}\int\limits_{-\infty}^{+\infty}\frac{\log\frac{\textup{Im}[\chi(\omega^{\prime})]}
{\lambda\omega^{\prime}\left|\chi(\omega^{\prime})\right|^2}}{\omega^{\prime}-\left(\omega+i\epsilon\right)}
\,d\omega^{\prime} \right]
\end{aligned}
\end{eqnarray}

Here we introduce the \emph{instantaneous spectrum at time $t$},
$E_{t}(\omega)$. It is the Fourier transform of $E_t^{-}$ (see the
lines following definition (\ref{Eq:uirrev})). We also introduce
the \emph{medium complexity factor} $\phi_+(\omega)$. Its
deviation from unity gives a measure of the effective
macroscopic/phenomenological incoherence of possible microscopic,
energy containing elements. Media for which $\phi_+(\omega)$ is
identically unity we call \emph{simple} media, the rest we call
\emph{complex}. In the case that the susceptibility $\chi$
corresponds to a single Lorentz oscillator medium, as in the case
considered by Loudon, $\phi_+(\omega)$ is identically one, and Eq.
(\ref{Eq:RI}) reduces to

\begin{eqnarray}
\begin{aligned}
u_{irrev.}[E](t) &=\lambda \int\limits_{-\infty}^{t} \dot
P^2(\tau) \, \, d\tau \label{Eq:RIL}
\end{aligned}
\end{eqnarray}
where, then, $\lambda$ is determined in terms of the
phenomenological damping parameter $\gamma$ and the effective
plasma frequency $\omega_p$:
\begin{eqnarray}
\begin{aligned}
\lambda &= \gamma/\omega_p^2.
\end{aligned}
\end{eqnarray}

In such case, then, and as claimed by Loudon, the relevant losses
correspond to the frictional losses generated by the single
Lorentz oscillator.

\section{Summary}

We have introduced notions of free energy and losses relevant to
the macroscopic behavior of passive, linear dielectric media.
These notions are relevant in the same regime in which the
macroscopic Maxwell equations are themselves relevant, i.e. in the
regime in which that theory specifies all measurable dynamics. In
particular, the macroscopic theory introduced is relevant to the
regimes in which the macroscopic, energy barrow-return process
describes the production of slow and fast light in passive,
linear, temporally dispersive media. Further communications will
describe the precise evolutions of the ideal medium-field
interactions giving rise to maximum energy recovery, i.e. those
"recovery" fields suggested by the variational definition
(\ref{Eq:uirrev}). The nature of the analogy of such recovery
fields in dissipative media with reversible processes in the
thermodynamic setting will be analyzed. Finally, the nonintuitive
evolution of the medium-field interaction on such recovery fields
in complex media will be exposed, e.g. the failure of monotonicity
in the evolution of the internal energy to its minimum value, even
on ideal recovery fields, and the identification of such as a
measure of the level of macroscopic disorder created by the
existence of many, competing degrees of freedom.

\begin{acknowledgments}
Conversations with Kurt Oughstun, Joseph Eberly, Peter Milonni,
Aephraim Steinberg, Raymond Chiao, and Michael Fleischhauer are
gratefully acknowledged. This research was supported in part by
the National Science Foundation under Grant No. PHY99-07949.
\end{acknowledgments}

\bibliography{PRL.bbl}

\end{document}